\documentclass[twocolumn,floatfix,superscriptaddress,amsmath,showkeys,aps,prb,eqsecnum]{revtex4-1}
\usepackage{bm}
\usepackage{xcolor}
\usepackage{hyperref}
\usepackage{makeidx}
\usepackage{amsmath}
\usepackage{graphicx}
\usepackage{dcolumn}
\usepackage{amsfonts}
\usepackage{amssymb}
\usepackage[latin9]{inputenc}
\usepackage{color}
\usepackage{epstopdf}

\begin{document}

\title{One-loop effective potential for two-dimensional competing scalar order parameters}

\author{Nei Lopes}
\email{nlsjunior12@gmail.com}
\affiliation{Centro Brasileiro de Pesquisas F\'{\i}sicas, Rua Dr. Xavier Sigaud 150, Urca, 22290-180, Rio de Janeiro , Brazil}
\author{Mucio A. Continentino}
\affiliation{Centro Brasileiro de Pesquisas F\'{\i}sicas, Rua Dr. Xavier Sigaud 150, Urca, 22290-180, Rio de Janeiro , Brazil}
\author{Daniel G. Barci}
\affiliation{Departamento de F\'{\i}sica Te\'orica, Universidade do Estado do Rio de Janeiro, Rua S\~ao Francisco Xavier 524, 20550-013,
Rio de Janeiro, RJ, Brazil.}

\date{\today}

\begin{abstract}
Using the method of the effective potential of quantum field theory, we compute the quantum corrections to the phase diagram of systems with competing order parameters. This is specially useful to study   {\it metallic systems} with competing antiferromagnetic and superconducting ground states. We focus on the two-dimensional (2d) case that is relevant for high T$_c$ superconductors and heavy fermion systems.  We consider two different types of couplings between the order parameters and obtain the modifications in the phase diagrams due to critical quantum fluctuations in these systems with conflicting orders. We consider $z=1$, as well as,  a dissipative $z=2$ dynamics,  typical of  antiferromagnetic metals close to the magnetic quantum critical point. Our results, when compared to those in the 3d case, show that these depend strongly on both dimensionality and dynamics of the propagators describing the excitations of the possible ordered states. We find  stable unconventional coexisting phases, as well as, the enhancement of the region of coexistence by fluctuations. These effects may be observed experimentally in many interesting cases of strongly correlated materials.

\end{abstract}

\maketitle

%%%%%%%%%%%%%%%%%%%%%%%%%%%%%%%%%%%%%%
\section{Introduction}
\label{sec:Introduction}
%%%%%%%%%%%%%%%%%%%%%%%%%%%%%%%%%%%%%%
Competing order parameters are a common feature in most strongly correlated materials~\cite{7,superaf,Tuson,Pagliuso,Chen,Sidorov,Chen2,Shermadini}. They produce very rich phase diagrams in terms of experimentally well controlled external parameters, such as, magnetic field, pressure and/or doping~\cite{sachdev,1.2}.

For instance, some pnictides~\cite{Raghu,Rafael2,Bang,Sunagawa,dalson,Shermadini} exhibit competing antiferromagnetic(AF)-superconductor(SC) orders, separated by a first-order phase transition~\cite{7,3,5,6}. SC-AF-Structural transitions, also appear in the iron-arsenide SC~\cite{3,4}. Coexistence between magnetic and SC orders are found in some iron-based compounds~\cite{Shermadini}. U and Ce-based heavy fermions materials~\cite{Pagliuso,Chen,Chen2} present coexistence of magnetism and superconductivity. Last but not least, magnetism and superconductivity are in close proximity in the  high-Tc cuprates~\cite{Pagliuso}. These experimental results are in contrast with the expected behavior of conventional SC, since magnetism and SC are usually competing phenomena\cite{Pagliuso} and there is a lot of ongoing effort to understand this kind of unusual behavior.

Nevertheless, competing orders can be traced back to the existence of competing states with different symmetries in the ground state of the system. The usual approach to describe the classical phase diagram near quantum criticality is the Landau-Ginzburg expansion of the free energy density in terms of the order parameters. At a microscopic level, the global phase diagram can be obtained by using mean-field approximations of  simplified model Hamiltonians. These procedures correctly capture those qualitative properties of the phase diagram that depend on symmetry of the problem.
 For example, in the case of two competing orders, time-reversal invariance leads to consider only  quartic couplings in the different fields. Examples are quadrupolar~\cite{7,8}  and spin nematic orderings~\cite{9,10}. Moreover, if both order parameters transform with the same irreducible representation of the symmetry group, a bilinear coupling is allowed~\cite{15,16}. In fact, this type of coupling is important to describe spin-density waves (SDW)~\cite{11,12}, orbital AF orders~\cite{13}, elastic instabilities of the atomic crystal lattice~\cite{14}, vortices in a multigap SC~\cite{14.1} and magnetic properties of the heavy fermion compound URu$_{2}$Si$_{2}$~\cite{15}.

However, at very low temperatures quantum fluctuations play an important role. A direct consequence is that not only the symmetry, but both dynamic and dimensionality of the system are relevant ingredients to determine the global topology of the phase diagram~\cite{sachdev,1.2}. In these cases, the Landau-Ginzburg approach is no longer appropriate.

In this paper we study the effects of quantum fluctuations on the  mean-field zero temperature phase diagram of systems with competing order parameters.  We are mainly interested in investigating two-dimensional  systems that comprise many interesting materials, like high T$_c$ cuprates,  heavy fermions compounds with tetragonal~\cite{Raghu,Rafael2,Bang,Sunagawa,dalson,Shermadini,Moore,Fujimori} structures and Fe-based systems, where AF and SC are in close proximity or coexist near a magnetic quantum critical point~\cite{1.2}. 

For simplicity, we consider two real scalar fields with biquadratic as well as bilinear interactions between both order parameters. Quantum fluctuations are computed by using the quantum field theory approach to  the one-loop effective potential~\cite{2,17,18}. 
Recently, in Ref. \onlinecite{nei-mucio-barci}, we have developed this method in detail and  applied it to a three-dimensional system in which the dynamics of the order parameters are driven by a Lorentz invariant~\cite{2,Mineevz1, Hertz, Mineevz2}, i.e., non-dissipative linear dispersion relation.  
We have found that the zero temperature bicritical point is robust to quantum fluctuations in the presence of biquadratic interactions. However, it becomes unstable under bilinear interactions, {\em i.e.,} quantum fluctuations induce a coexistence phase.

In general, we expect that lower dimensionalities enhance the effect of quantum fluctuations. With this motivation, we extend the results of Ref.~\onlinecite{nei-mucio-barci} to bidimensional systems considering two different types of dynamics for the order parameters: non-dissipative linear dispersion ($z=1$)~\cite{2,Mineevz1, Hertz, Mineevz2}, which can be related to the case of interacting magnetic excitons, as well as dissipative quadratic dispersion relations ($z=2$)~\cite{2, Mineevz1, Hertz, Mineevz2} that are normally associated with paramagnons in itinerant AF or dissipative modes in SC systems near their respective  quantum critical points.

We find that, for $d=2$ and $z=1$, the bicritical point is unstable to any finite (\textit{positive}) coupling, bilinear or biquadratic, thus  implying  that quantum fluctuations lead to stable phase coexistence. On the other hand, in the coexistence region our results show that quantum corrections in the presence of both couplings tend to further increase this region. Moreover, we obtain that under well-defined conditions, there may be a region in the phase diagram where the coexisting phases are metastable giving rise to non-homogeneous ground states. This behavior should be compared with the three-dimensional case~\cite{nei-mucio-barci} of  cubic structure compounds. There, biquadratic interactions did not modify the bicritical point, while a bilinear  coupling gave rise to coexistence~\cite{nei-mucio-barci}. In  the coexistence region, the fluctuations associated with these couplings were in competition and produced opposite effects~\cite{nei-mucio-barci}.  

In contrast, for $d=2$ and $z=2$,  where only a  bi-quadratic coupling is allowed by symmetry, the bicritical point survives to quantum fluctuations. Also, we find that these interactions tend to enhance the coexistence region. For $d=3$ and $z=2$, such that,  $d_{eff}=2+3>d_c=4$, with $d_c$ the \textit{upper critical dimension}, we confirm that quantum fluctuations do not give rise to qualitative changes in the classical phase diagram, as expected.

The paper is organized as follows: 
in section \ref{sec:pd}, we describe our model and the method used to compute the quantum corrections to the zero temperature phase diagrams. In \S \ref{sec:Quantum} we present the effective potential for two-dimensional systems in the cases of $z=1$ and $z=2$ dynamics, presenting  drastic changes on the classical phase diagram when we take into account quantum fluctuations. Finally, in section \ref{sec:Conclusions} we summarize and discuss the main results.
  
%%%%%%%%%%%%%%%%%%%%%%%%%%%%%%%
\section{Zero temperature phase diagram for two interacting order parameters}
\label{sec:pd}
%%%%%%%%%%%%%%%%%%%%%%%%%%%%%%%%%
We consider two real scalar fields $\phi_1$ and $\phi_2$, satisfying a usual $\phi^4$ Landau expansion, invariant under the discrete group 
$Z_2\times Z_2$. The simplest interaction that preserves this symmetry is biquadratic, $\phi_1^2\phi_2^2$, since this term is invariant under the transformation $\phi_1\to -\phi_1$ {\em or} $\phi_2\to -\phi_2$ independently. Moreover, since both fields transforms with the same representation of the symmetry group, we can also consider a bilinear coupling of the form 
$\phi_1\phi_2$ that breaks the original symmetry to $Z_2\times Z_2\to Z_2$, {\em i.e.,} with this interaction, the theory is invariant under simultaneous sign changes of the fields. In magnetic systems, this bilinear term is related to time-reversal symmetry breaking~\cite{15,11}.  
In the presence of a bilinear interaction, it is convenient to work in a rotated~\cite{nei-mucio-barci} bases $(\phi_1,\phi_2)\to (\varphi_1,\varphi_2)$, in order to diagonalize the quadratic part of the Landau expansion of the free energy density.  In this bases the free energy density reads, 
\begin{equation}
V_{cl}(\varphi_i)= V_{cl}^{(2)}(\varphi_i)+ V_{cl}^I(\varphi_i) \; , 
\label{eq:Vcl}
\end{equation}
where $\varphi_i$ represent either $\varphi_1$ or $\varphi_2$. 
The quadratic contributions are
\begin{equation}
V_{cl}^{(2)}(\varphi_i)= r_{1}(P)\varphi_{1}^{2}+r_{2}(P)\varphi_{2}^{2}\; , 
\label{eq:Vcl2}
\end{equation}
while the interaction terms read
\begin{equation}
V^I_{cl}(\varphi_i)=\lambda_{1}\varphi_{1}^{4}+\lambda_{2}\varphi_{2}^{4}+\lambda_{12} \varphi_{1}^{2} \varphi_{2}^{2}+\delta_{1} \varphi_{1}^{3}\varphi_{2}+\delta_{2} \varphi_{1}\varphi_{2}^{3} \; .
\label{eq:VclI}
\end{equation}
Eq.~(\ref{eq:Vcl}), together with Eqs.~(\ref{eq:Vcl2}) and (\ref{eq:VclI}), represent the most general quartic free energy density that can be built with two real scalar order parameters, keeping a global $Z_2$ invariance. The originally bilinear term $\phi_1\phi_2$ (that breaks $Z_2\times Z_2$) has now been shifted to the quartic terms $\varphi_{1}\varphi_{2}^{3}$ and $\varphi_{1}^{3}\varphi_{2}$.  The coefficients are, in principle, arbitrary constants.
This free energy density leads to  different regimes.  For example, at $r_1(P)=r_2(P)=0$, which are functions of a external control parameter $P$, as pressure, there is a quantum bicritical point for both $\delta_{1,2}$ positive. Moreover, depending on the specific values of the parameters, $V_{cl}$  also supports coexistence regions\cite{7,10,15,nei-mucio-barci}. The detailed classical phase diagram is computed by solving 
$\partial V_{cl}/\partial\varphi_i=0$. 

Quantum corrections to the mean-field phase diagram can be obtained by means of the effective potential\cite{Rudnei}
\begin{equation}
V_{eff}(\varphi_i)=V_{cl}(\varphi_i)+\hbar\;  \Gamma^{(1)}(\varphi_i) + O(\hbar ^2)\; , 
\label{eq:Veff}
\end{equation}
where $V_{cl}(\varphi_i)$ is the classical potential given by Eq.~(\ref{eq:Vcl}) and $\Gamma^{(1)}(\varphi_i)$ contains quantum fluctuations at first order in an $\hbar$ expansion (\textit{one-loop} expansion). The actual phases of the system are reached by computing $\partial V_{eff}/\partial\varphi_i=0$~\cite{nei-mucio-barci, Rudnei}.
The one-loop quantum corrections are given by ~\cite{2,17,18,nei-mucio-barci},
\begin{equation}
\Gamma^{(1)}(\varphi_i)=\frac{1}{2}\int \frac{d^{d}k}{(2\pi)^{d}} \ln \left(\det [1-M(k)]\right)+ \mbox{counterterms}.
\label{eq:Gamma}
\end{equation}
$M(k)$ is a matrix whose  elements $[M]_{lm}$ are given by
\begin{equation}
[M]_{lm}=-G_{0}^{(l)}(k)\left[\frac{\partial^{2}V^{I}_{cl}({{\varphi_{i}}})}{\partial \varphi_{l}\partial \varphi_{m}}\right]_{{\varphi_{i}=\varphi_{c}}}
\; .
\label{eq:M}
\end{equation}
Here, $G_{0}^{l}(k)$ are the propagators of the $\varphi_l$ fields and $V^{I}_{cl}$ is given by Eq. (\ref{eq:VclI}). The counterterms in Eq. (\ref{eq:Gamma}) are introduced to renormalize the theory, making the observables \textit{cut-off} independent. 

Dynamical effects are included through the frequency dependent propagators~\cite{nei-mucio-barci,Mineevz1,Hertz,Mineevz2}.
We will consider two different types of propagators in Euclidean \textit{momentum} space.
\begin{equation}
G_{0}^{(1,2)}(\omega,\vec{q})=\frac{1}{k^{2}+r_{1,2}},
\label{eq:Gz1}
\end{equation}
where $k^{2}=\omega^{2}+q^{2}$, characterizes a Lorentz invariant critical theory, i.e.,   with a linear dispersion implying a dynamical exponent $z=1$.  This kind of exponent is usually related to interacting magnetic excitons, where the effective dimensionality is simply increased by 1~\cite{Hertz}.
We will also be interested in  the case of $z=2$, given by the propagator
\begin{equation}
G_{0}^{(1,2)}(\omega,\vec{q})=\frac{1}{\tau\left|\omega\right|+q^{2}+r_{1,2}} \; ,
\label{eq:Gz2}
\end{equation}
that belongs to a different class of systems characterized by quadratic dispersion relations, which are normally associated to paramagnons in itinerant AF\cite{Hertz} or overdamped modes in SC\cite{1.2} near their quantum critical points.

With these two different types of dynamics we cover the most interesting cases of competing orders in strongly correlated materials. In the next section we compare the  results of $V_{eff}$, for dimensions $d=2$ and $d=3$, with dynamics $z=1$ and $z=2$.

%%%%%%%%%%%%%%%%%%%%%%%%%%%%%%%%%%%
\section{Quantum corrections: one-loop effective potential}
\label{sec:Quantum}
%%%%%%%%%%%%%%%%%%%%%%%%%%%%%%%%%%%%
In this section we compute and compare the effective potential $V_{eff}$ given by Eq.~(\ref{eq:Veff}) for both cases of different, spatial dimensions and  dynamics. 
We closely follow the procedure described in Ref. \onlinecite{nei-mucio-barci}, where we have explicitly calculated the effective potential in a three-dimensional system with dynamical exponent $z=1$. In that reference, we have analyzed two cases. The case of a quantum bicritical point, for which $\varphi_1=\varphi_2=0$, and the case when, at classical level, there is a coexistence region with  $\varphi_1\neq 0$ and $\varphi_2\neq 0$ in the ground state.  

We have studied the effect of the quartic interactions, parametrized by the coupling constants $\lambda_{12}$ and $\delta_{1,2}$, on the phase diagram in both cases. The main results were the following:  quantum fluctuations do not modify the phase diagram near the quantum bicritical point, when considering inter-species interactions of the form $\lambda_{1,2}\neq 0$ and $\delta_{1,2}=0$. However, fluctuations in the presence of a bilinear interactions $\delta_{1,2}\neq 0$, spontaneously breaks $Z_2$ symmetry, favoring a coexistence phase. In that way, quantum fluctuations deeply change the classical phase diagram. 
On the other hand, in the coexistence region of the phase diagram, we have found that, due to the competition between the order parameters, 
$\lambda_{12}$ tends to shrink the coexistence region, while $\delta_{1,2}$ increases that region, consistent with the symmetry breaking produced at the bicrictical point. 

In three dimensions, we do not expect to have a different behavior when considering dissipation like dynamics with $z=2$. The reason is that, in this case, the effective dimension,  $d_{eff}=3+2$, is greater that the \textit{upper critical dimension} $d_c=4$~\cite{sachdev,1.2}. In this case, we expect that quantum fluctuations do not qualitatively modify the mean-field result. In fact, we have explicitly computed the effective potential in this case, and we have confirmed that the bicritical point survives fluctuation effects. 

%%%%%%%%%%%%%%
\subsection{Two-dimensional system with linear dispersion relation}
Here, we extend the calculations presented in Ref. \onlinecite{nei-mucio-barci} to bidimensional systems. The effective dimension in this case is 
$d_{eff}=3$, less than the upper critical dimension. Then, we expect that quantum fluctuations will play an important role.  It is clear that, in this case, the effective potential technique is not adequate to compute critical exponents. However, it captures the global features of the phase diagram and in particular the stability of the critical point. 

\subsubsection{Bicritical point}
At mean-field level, the system of Eq.~(\ref{eq:Vcl}) has a bicritical point for $r_{1}(P)=r_{2}(P)=0$ and $\delta_{i}$ positive. To compute quantum fluctuations, we fix these values and evaluate the integrals in Eq.~(\ref{eq:Gamma}).  As usual, since the \textit{momentum} integrals are ultraviolet divergent, we regularize  them with a \textit{momentum} (and frequency) \textit{cut-off} and  we add counterterms to eliminate the dependence of the effective potential on the \textit{cut-off}. From now on, we consider $\hbar=1$ in Eq.~(\ref{eq:Veff}). Thus, we obtain
\begin{align}
\Gamma^{(1)}(\varphi_1,\varphi_2)=&-\frac{\sqrt{2}}{(2 \pi)}\bigg[\frac{1}{3}\left(b_{1}^{3/2}+b_{2}^{3/2}\right)+ \nonumber \\
&+\frac{\left(3\delta_{1}\varphi_{1}^{2}+3\delta_{2}\varphi_{2}^{2}+4\lambda_{12}\varphi_{1}\varphi_{2}\right)^{2}}{2(\sqrt{b_{2}}+\sqrt{b_{1}})}\bigg]
\label{eq: Gammad2z1}
\end{align}
where,
\begin{align}
b_{1}&=12\lambda_{1}\varphi_{1}^{2}+2\lambda_{12}\varphi_{2}^{2}+6\delta_{1}\varphi_{1}\varphi_{2} \; ,
\label{eq:b1} \\
b_{2}&=12\lambda_{2}\varphi_{2}^{2}+2\lambda_{12}\varphi_{1}^{2}+6\delta_{2}\varphi_{1}\varphi_{2} \; .
\label{eq:b2}
\end{align}
The first important observation is that since $\Gamma^{(1)}<0$,  $\varphi_c=0$ is no longer a minimum of the effective potential 
$V_{eff}$.
\begin{figure}[!ht]
\begin{center}
\includegraphics[scale=0.65]{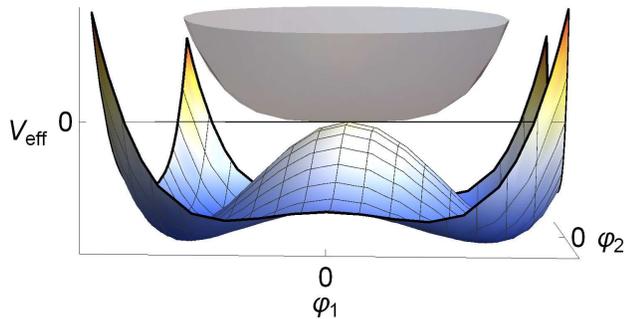} 
\end{center}
\caption{(Color online) The effective potential for a bicritical point with  (square lines) and without  (gray smooth) quantum corrections. Quantum corrections induce coexistence, i.e., $\varphi_c=0$ is no longer a minimum of the effective potential (see text).}
\label{Fig.1}
\end{figure}
In fact, $V_{eff}$ has a minimum for $\varphi_c \neq 0$, signaling a spontaneous symmetry breaking, as can be seen in Fig.~\ref{Fig.1}. 

\begin{figure}[!ht]
\begin{center}
\includegraphics[scale=0.9]{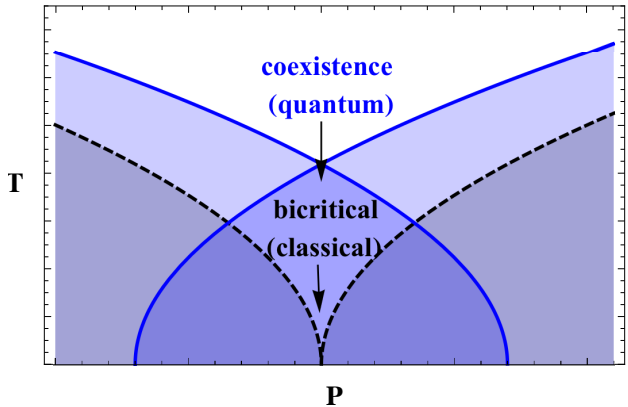} 
\end{center}
\caption{(Color online) Schematic phase diagram showing temperature as a function of pressure for bicritical point with (blue full line) and without (black dashed line) quantum corrections. Quantum corrections give rise to coexistence (see text).}
\label{Fig.2}
\end{figure}

Thus, the bicritical point is unstable and quantum fluctuations induce coexistence (see Fig.~\ref{Fig.2}). 
Our One-loop effective potential results can be compared with a perturbative renormalization group analysis. In ref. \onlinecite{Wang}, a very detailed  Wilson renormalization group scheme was implemented at one-loop order for an equivalent model (without linear couplings). Their results show that the Gaussian fix point is unstable with all coupling constants growing unbounded very quickly, signaling the absence of non-trivial critical points.  These results are in complete agreement with ours.

Interestingly, this dramatic change in the mean-field phase diagram occurs for any finite (\textit{positive}) value of the couplings $\lambda_{12}$ as well as $\delta_{1,2}$. In other words, quantum fluctuations provide stability for the coexistence of phases, whatever the coupling. This result is quite different from its three-dimensional version, where the mean-field bicritical point is unstable under bilinear interactions, proportional to $\delta_{1,2}$. However, it is stable under the biquadratic interaction $\lambda_{12}$.

\subsubsection{Coexistence region}
Depending on the parameters of the theory, the mean-field phase diagram supports coexistence regions\cite{nei-mucio-barci}. In particular, coexistence is possible whether $\lambda_{12}^2<4\lambda_1 \lambda_2+ \delta_2 \lambda_1 + \delta_1 \lambda_2 +\delta_1 \delta_2 /4$, with $\delta_{1,2} \le 0$.

To compute quantum fluctuations in a coexistence phase,  we will focus in one of the quantum critical points, say $r_1=0$.  At this point, 
$r_2<0$ and $\varphi_2\neq 0$ since this is an ordered phase. The calculation in the opposite region, $r_2=0$ with $r_1<0$, is completely analogous.   

Computing the integrals of Eq.~(\ref{eq:Gamma}), and proceeding with the renormalization process to eliminate the \textit{cut-off} dependence, we obtain for the quantum corrections, 
\begin{align}
\Gamma^{(1)}(\varphi_1,\varphi_2)=&-\frac{\sqrt{2}}{(2 \pi)}\bigg[\frac{1}{3}\left[b_{1}^{3/2}+\left(b_{2}-\left|r_{2}\right|\right)^{3/2}\right]+ \nonumber \\
&+\frac{\left(3\delta_{1}\varphi_{1}^{2}+3\delta_{2}\varphi_{2}^{2}+4\lambda_{12}\varphi_{1}\varphi_{2}\right)^{2}}{2(\sqrt{r_{2}+b_{2}}+\sqrt{b_{1}})}\bigg]
\label{eq: num46.1}
\end{align}
where $b_{1,2}$ are given by Eqs.~(\ref{eq:b1}) and (\ref{eq:b2}).

\begin{figure}[!ht]
\begin{center}
\includegraphics[scale=0.9]{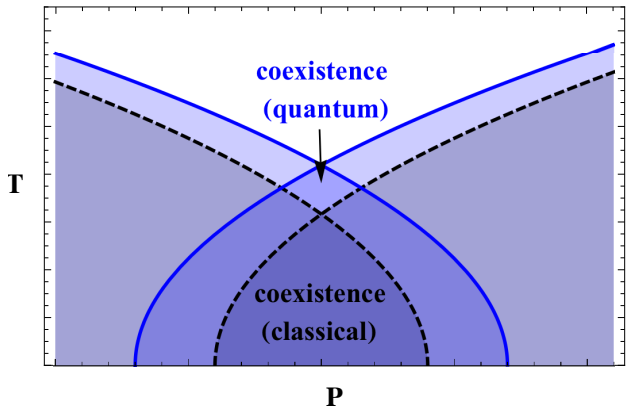} 
\end{center}
\caption{(Color online) Schematic phase diagram showing temperature as a function of pressure for coexistence region with (blue full line) and without (black dashed line) quantum corrections. Quantum corrections enhance coexistence region (see text).}
\label{Fig.3}
\end{figure}

With this correction, the effective potential $V_{eff}$ will continue to have minima for $\varphi_{1,2}\neq 0$. Note that this correction is quite different from the bicritical case, due to the appearance of the \textit{mass term} contribution ($\left|r_{2}\right|$). On the other hand, from the $(b_{2}-\left|r_2\right|)^{3/2}$ term, it is clear that $|r_2|<b_2$. If this condition is not satisfied, the effective potential gets an imaginary part, signaling that the homogeneous coexistence is metastable\cite{Weinberg}. In this case, the ground state is no longer homogeneous, giving rise to domain formation. In the region of stability, $|r_2|<b_2$, both $\lambda_{12}$ and $\delta_{1,2}$ tends to enhance the coexistence region (see Fig.~\ref{Fig.3}), which is consistent with the fact that in the bicritical phase, both interactions produce a symmetry breaking, tending to order both phases, thus ensuring the stability of the coexistence region.
However, this is in contrast to the three-dimensional case, where $\lambda_{12}$ does not increase any order; in fact, it tends to shrink the coexistence region due to simple competition already present at mean-field level.

%%%%%%%%%%%%%
\subsection{Two-dimensional system with dissipative quadratic dispersion relation}
In this subsection we compute the effective potential of a two-dimensional system given by Eq.~(\ref{eq:Vcl}) with propagators characterized  by the same dynamic exponent $z=2$, Eq.~(\ref{eq:Gz2}), which is usually associated to dissipative modes in SC or   paramagnons in itinerant AF~\cite{2,Mineevz1, Hertz, Mineevz2,nei-mucio-barci} near their respective quantum critical points.
Indeed, we are interested in dealing with SC and AFM competing orders, which are well described by this kind of dynamics~\cite{2,1.2}. Consider for instance the Landau expansion of the free energy density for an AFM order parameter $\varphi_1$ coupled with a complex SC order parameter $\Delta$,
\begin{equation}
V_{cl}=r_1\varphi_{1}^{2}+r_2|\Delta|^2+ \lambda_1 \varphi_1^4+\lambda_2|\Delta|^4 + \lambda_{12}\varphi_1^2|\Delta|^2\; .
\label{eq:VAFMSC}
\end{equation}

The SC order parameter can be parametrized as  $\Delta=\varphi_2+ i \varphi_3$ or, equivalently, $\Delta=|\Delta| e^{i\theta}$.
The energy density is invariant under transformations of the group $U(1)\times Z_2$, where the $U(1)$ group is related with the phase transformations, $\theta\to\theta+\delta\theta$, and $Z_2$ denote sign changes of the real scalar order parameter $\varphi_1$. Since each order parameter transform with a different symmetry group, bilinear couplings are forbidden.

In the superconductor ordered phase, $\Delta\neq 0$,  there is one massless Goldstone  mode associated with phase fluctuations $\delta\theta$. In two spatial dimensions at finite temperature, the $\theta(x)$ correlation function  diverges logarithmically with the size of the sample,  completely disordering the system. This is nothing but the Mermin-Wagner theorem~\cite{mermin-wagner} that states that  a continuous symmetry cannot be spontaneously broken at finite temperature in $2$D. However, at zero temperature, time adds an extra dimension in the problem and the $\theta$ correlation function is infrared finite.  At one loop order,  phase fluctuations decouple from the longitudinal ones and can be absorbed in a global normalization constant.  Thus, in this case,  the Goldstone mode cannot qualitatively change the character of the  quantum phase transition. This behavior is quite different  in the presence of a magnetic field, since $\theta(x)$ couples with the vector potential, producing the Meissner effect.

Thus, we can safely choose in Eq.~(\ref{eq:VAFMSC}) a particular direction of the SC order parameter, for instance, $\varphi_{3}=0$, and compute the effect of the longitudinal fluctuations, $\varphi_2$.  In these circumstances, the classical energy density is completely equivalent to Eq.~(\ref{eq:Vcl}) with $\delta_{1,2}=0$ and the formal calculation of the effective potential $V_{eff}$ follows the same lines of the previous cases.

\subsubsection{Bicritical point}
The system at mean-field level has a bicritical point at $r_1(P)=r_2(P)=0$, this result does not depend neither on dimensionality nor on the dynamics. 
Thus, we fix these values in Eq.~(\ref{eq:Vcl}) and compute $\Gamma^{(1)}$, Eq.~(\ref{eq:Gamma}), using the $z=2$ propagators, Eq.~(\ref{eq:Gz2}). Again, we regularize the integrals with an ultraviolet \textit{cut-off}, and renormalize the effective potential by using the proper counterterms. We obtain, 
\begin{align}
V_{eff}(\varphi_1,\varphi_2)&=\lambda_1'\varphi_{1}^{4}+\lambda_2'\varphi_{2}^{4}+\lambda_{12}'\varphi_{1}^{2}\varphi_{2}^{2} \ +
\label{eq:Veffz2bicritical}  \\
&+\frac{1}{(2 \pi)^{2}}\left[8\lambda_{12}^{2}\varphi_{1}^{2}\varphi_{2}^{2}\left(\frac{b_{1}\ln(b_{1})-b_{2}\ln(b_{2})}{b_{1}-b_{2}}\right)\right]
\nonumber 
\end{align}
where $b_{1,2}$ are given by Eqs. (\ref{eq:b1}) and (\ref{eq:b2}) with $\delta_{1,2}=0$,
and the $prime$ quantities represent renormalized effective couplings.

\begin{figure}[!ht]
\begin{center}
\includegraphics[scale=0.7]{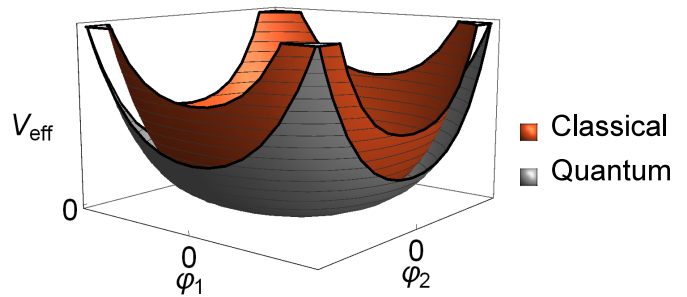} 
\end{center}
\caption{(Color online) The effective potential for a bicritical point with (gray (external plot)) and without (orange (internal plot)) quantum corrections. The only minimum of $V_{eff}$ remains at origin (see text).}
\label{Fig.4}
\end{figure}

The important result is that the only minimum of $V_{eff}$, computed from $\partial V_{eff}/\partial\varphi_i=0$, is $\varphi_c=0$, as can be seen in Fig.~\ref{Fig.4}.
Then, the bicritical point is robust and survives at quantum level. In other words, there are no qualitative changes in the mean-field phase diagram.
For consistency reasons, it is very simple to check that the effective potential of Eq.(\ref{eq:Veffz2bicritical}) satisfies the Callan-Symanzik renomalization group equation\cite{Meissner}, with $\beta$ and $\gamma$ functions signaling a stable fixed point. 
This result is in contrast with that obtained with $z=1$ dynamics, discussed in the previous section.

\subsubsection{Coexistence region}
For completeness, we compute quantum corrections in the region of  coexistence of dissipative systems ($z=2$).
We focus in the neighbourhood of one of the quantum critical points, say $r_2=0$ (but $r_1<0$).
We compute the $\Gamma^{(1)}$, using Eq. (\ref{eq:Gamma}) with the same procedure of the previous cases. We obtain
 \begin{align}
V_{eff}&=r_1\varphi_{1}^{2}+\lambda'_1\varphi_{1}^{4}+\lambda'_2\varphi_{2}^{4}+\lambda'_{12}\varphi_{1}^{2}\varphi_{2}^{2} \ +
\label{eq:Veffcoexistencez2} \\
&+\left(\frac{2}{\pi}\right)^2\left[\left|r_1\right|\left(\lambda_{12}\varphi_{2}^{2}+6\lambda_1\varphi_{1}^{2}\right)\ln\left(\frac{b_{1}+2r_1}{2r_1}\right)\right]\; , 
\nonumber
\end{align}
where, as before, $b_{1}$ is given by Eq. (\ref{eq:b1}) with $\delta_{1,2}=0$,
and the $prime$ quantities represent renormalized effective couplings.
The important observation here is that the contribution of quantum fluctuations is {\em negative}. This comes out because the argument of the 
$\ln$ function  is always less than one. On the other hand, we are considering the coupling $\lambda_{12}>0$. Therefore, 
the interaction between both order parameters {\em enhance the region of coexistence}, very similar to Fig.~\ref{Fig.3}, despite that  classically, $\lambda_{12}$ tends to shrink this region, i.e., quantum fluctuations provide stability for the coexistence of different phases. Once again, this is in contrast with the $d=3$ non-dissipative case ($z=1$), where $\lambda_{12}$ tries to frustrate coexistence.   

%%%%%%%%%%%%%%%%%%%%%%%%%%%%%%%%%%
\section{Summary and conclusions} 
\label{sec:Conclusions}
%%%%%%%%%%%%%%%%%%%%%%%%%%%%%%%%%%
In this paper, we have addressed the effect of quantum fluctuations on the zero temperature classical phase diagram of two competing order parameters. In particular, we have evaluated the one-loop effective potential of two real scalar order parameters, interacting via biquadratic as well as bilinear couplings. While at mean-field level, the phase diagram only depends on symmetry, at a quantum level, dimensionality and dynamics are essential ingredients to ensure the stability of the coexistence of phases. 

For this reason, we have investigated the effective potential both in three and two spatial dimensions, considering dynamics characterized by linear and quadratic dispersion relations, i.e., by dynamic critical exponents, $z=1$ and $z=2$. These possibilities cover most of the interesting cases of competing phases observed in strongly correlated materials. Our results provide  requirements for the emergence of stable, unconventional coexisting phases, as observed experimentally in Fe-based~\cite{Raghu,Rafael2,Bang,Sunagawa,dalson,Shermadini} SC, U and Ce-based~\cite{Pagliuso,Chen,Chen2,Moore,Fujimori} heavy fermions compounds, or even in high-T$_c$ cuprates~\cite{Pagliuso}, through the effects of quantum fluctuations.

Two-dimensional quantum fluctuations characteristic of compounds with tetragonal structures  are stronger than three-dimensional ones, producing more drastic effects on the classical phase diagram, as expected. We have found that for non-dissipative dynamics ($z=1$), related to interacting magnetic excitons, where the effective dimensionality is simply increased by 1, the bicritical point is unstable and quantum fluctuations induce phase coexistence for both types of interactions between order parameters. This is a purely quantum effect and resembles the physics of the Coleman-Weinberg mechanism~\cite{18},  where coupling to a gauge field gives rise to spontaneous symmetry breaking. 

These effects are in contrast with the three-dimensional case where, although the bilinear interaction spontaneously breaks global $Z_2$ symmetry, the biquadratic interaction does not modify the bicritical point.  
On the other hand, the bicritical point is robust and survives at quantum level, in $d=2$ as well as in $d=3$, in the presence of  dissipative dynamics ($z=2$). This is the case of itinerant AF\cite{Hertz} or SC\cite{1.2} with dissipative modes with quadratic dispersion near their quantum phase transitions. The latter is in agreement with the expectation that for effective dimensions $d_{eff}=d+z > 4$, we should not expect drastic changes of the mean-field results.

With respect to the region of coexistence of both order parameters, we found that two-dimensional quantum fluctuations enhance the coexistence region whatever  is the dynamics. Thus, quantum effects change the mean-field tendency of the biquadratic interaction to shrink this region, i.e., they provide stability for the coexistence of different  phases.
Moreover, we obtain that, for bidimensional system with linear dispersion relation, there may be a sector of the phase diagram, under well defined conditions, where the coexistence region is metastable, favoring domain formation and a non-homogeneous ground state, which may be observed experimentally. 

In summary, we have shown the importance of considering the effect of quantum fluctuations  on the mean-field phase diagrams of  systems with competing order parameters. The effect of these quantum corrections is essential to understand the emergence of stable unconventional coexisting orders, experimentally observed  in strongly correlated materials. Our results show how symmetry,  dynamic and dimensionality determine the  nature of the phase diagrams.

%%%%%%%%%%
\acknowledgments
The Brazilian agencies {\em Conselho Nacional de Desenvolvimento Cient\'\i fico e Tecnol\'ogico} (CNPq), {\em Funda\c c\~ao  Carlos Chagas Filho de Amparo \`a Pesquisa do Estado do Rio de Janeiro} (FAPERJ), and {\em Coordena\c c\~ao de Aperfei\c coamento de Pessoal de N\'\i vel Superior} (CAPES) are acknowledged for partial financial support.


\begin{thebibliography}{99}
\bibitem{7} R.M. Fernandes and J. Schmalian, Phys. Rev. {\bf B 82}, 014521 (2010).
\bibitem{superaf} R. Movshovich, T. Graf, D. Mandrus, J. D. Thompson, J. L. Smith, and Z. Fisk, Phys. Rev. {\bf B 53}, 8241 (1996); N.  D. Matur, F. M. Grosche, S. R. Julian, I. R. Walker, D. M. Freye, R. K. W. Haselwimmer and G. G. Lonzarich,  Nature {\bf 394}, 39 (1998).
\bibitem{Tuson} T. Park et al, Nature Letters {\bf 440} (2006).
\bibitem{Pagliuso} P.G. Pagliuso, C. Petrovic, R. Movshovich, D. Hall, M.F. Hundley,J.L. Sarrao, J.D. Thompson and Z. Fisk, Phys. Rev. B, {\bf 64}, 100503(R) (2001).
\bibitem{Chen} K. Chen, F. Strigari, M. Sundermann, Z. Hu, Z. Fisk, E.D. Bauer, P.F.S. Rosa, J.L. Sarrao, J.D. Thompson, J. Herrero-Martin, E. Pellegrin, D. Betto, K. Kummer, A. Tanaka, S. Wirth, and A. Severing, Phys. Rev. B, {\bf 97}, 045134 (2018).
\bibitem{Sidorov} V.A. Sidorov, M. Nicklas, P.G. Pagliuso, J.L. Sarrao, Y. Bang, A.V. Balatsky and J.D. Thompson, Phys. Rev. Lett., {\bf 89}, 157004 (2002).
\bibitem{Chen2} G.F. Chen, K.Matsubayashi, S. Ban, K. Deguchi and N.K. Sato, Phys. Rev. Lett., {\bf 97}, 017005 (2006).
\bibitem{Shermadini} Z. Shermadini, A. Krzton-Maziopa, M. Bendele, R. Khasanov, H. Luetkens, K. Conder, E. Pomjakushina, S. Weyeneth, V. Pomjakushin, O. Bossen and A. Amato, Phys. Rev. Lett., {\bf 106}, 117602 (2011).
\bibitem{sachdev} S. Sachdev, \textit{Quantum Phase Transitions}, Cambridge University Press, UK (1999).
\bibitem{1.2} M.A. Continentino, {\it Quantum scaling in many-body systems: an approach to quantum phase transitions}, Cambridge University Press, (2017).
\bibitem{Raghu} S. Raghu, X.L. Qi, C.X. Liu, D.J. Scalapino and S.C. Zhang, Phys. Rev. B , {\bf 77}, 220503(R) (2008).
\bibitem{Rafael2} R.M. Fernandes and A.V. Chubukov,  Rep. Prog. Phys., {\bf 80}, 014503 (2017).
\bibitem{Bang} Y. Bang and G.R. Stewart,  J. Phys.: Condens. Matter, {\bf 29}, 123003 (2017).
\bibitem{Sunagawa} M. Sunagawa, et al., Nature srep 04381 (2014).
\bibitem{dalson} D. E. Almeida, R. M. Fernandes, and E. Miranda, Phys. Rev. {\bf B  96}, 014514 (2017).
\bibitem{3} H. Luetkens, H.-H. Klauss, M. Kraken, F. J. Litterst, T. Dellmann, R. Klingeler, C. Hess, R. Khasanov, A. Amato, C. Baines,
M. Kosmala, O. J. Schumann, M. Braden, J. Hamann-Borrero, N. Leps, A. Kondrat, G. Behr, J. Werner, and B. B\"uchner, Nature Mater. {\bf 8}, 305 (2009).
\bibitem{5} C. R. Rotundu, D. T. Keane, B. Freelon, S. D. Wilson, A. Kim, P. N. Valdivia, E. Bourret-Courchesne, and R. J. Birgeneau, Phys. Rev. {\bf B 80}, 144517 (2009).
\bibitem{6}T. Goko, A. A. Aczel, E. Baggio-Saitovitch, S. L. Budko, P. C Canfield, J. P. Carlo, G. F. Chen, P. Dai, A. C. Hamann, W. Z. Hu, H. Kageyama, G. M. Luke, J. L. Luo, B. Nachumi, N. Ni, D. Reznik, D. R. Sanchez-Candela, A. T. Savici, K. J. Sikes, N. L. Wang, C. R. Wiebe, T. J. Williams, T. Yamamoto, W. Yu, and Y. J. Uemura, Phys. Rev. {\bf B 80}, 024508 (2009).
\bibitem{4} A. J. Drew, Ch. Niedermayer, P. J. Baker, F. L. Pratt, S. J. Blundell, T. Lancaster, R. H. Liu, G. Wu, X. H. Chen, I. Watanabe, V.
K. Malik, A. Dubroka, M. Rassle, K. W. Kim, C. Baines, and C. Bernhard, Nature Mater. {\bf 8}, 310 (2009).
\bibitem{8} P. Santini and G. Amoretti, Phys. Rev. Lett. {\bf 73}, 1027 (1994).
\bibitem{9} V. Barzykin and L. P. Gorkov, Phys. Rev. Lett. {\bf 70}, 2479 (1993).
\bibitem{10} D.G. Barci, R.V. Clarim and  N.L. Silva Junior, Phys. Rev. {\bf B 94}, 184507 (2016).
\bibitem{15} V. P. Mineev and M. E. Zhitomirsky, Phys. Rev. {\bf B 72}, 014432 (2005).
\bibitem{16} D. F. Agterberg and M. B. Walker, Phys. Rev. {\bf B 50}, 563 (1994).
\bibitem{Moore} D. P. Moore et al, Physica B, 312 \& 313, 134 (2002).
\bibitem{Fujimori} S. Fujimori et al, Journal of the Physical Society of Japan, \textbf{81}, 014703 (2012).
\bibitem{11} A. P. Ramirez, P. Coleman, P. Chandra, E. Br\"uck, A. A. Menovsky, Z. Fisk, and E. Bucher, Phys. Rev. Lett. {\bf 68}, 2680 (1992).
\bibitem{12} H. Ikeda and Y. Ohashi, Phys. Rev. Lett. {\bf 81}, 3723 (1998).
\bibitem{13} P. Chandra, P. Coleman, J. A. Mydosh, and V. Tripathi, Nature (London) {\bf 417}, 831 (2002); P. Chandra, P. Coleman, and J. A.
Mydosh, Physica {\bf B} 312313, 397 (2002).
\bibitem{14} M. Zacharias, I. Paul and M. Garst, Phys. Rev. Lett. {\bf 115}, 025703 (2015).
\bibitem{14.1} M. E. Zhitomirsky and V.-H. Dao, Phys. Rev. {\bf B 69}, 054508 (2004).
\bibitem{2} A.S. Ferreira, M.A. Continentino and E.C. Marino, Phys. Rev. {\bf B 70}, 174507 (2004).
\bibitem{17} G. Jona-Lasinio, Nuovo Cimento {\bf 34}, 1790 (1964).
\bibitem{18} S. Coleman and E. Weinberg, Phys. Rev. {\bf D 7}, 1888 (1973).
\bibitem{nei-mucio-barci} N.L. Silva Jr, M.A. Continentino and D.G. Barci, J. Phys.: Condens. Matter, {\bf 30} 255402 (2018).
\bibitem{Mineevz1} V.P. Mineev, Pisma Zh. \'Eksp. Teor Fiz. {\bf 66}, 655 (1997)[JETP Lett. 66, 693 (1997)].
\bibitem{Hertz} J.A. Hertz, Phys. Rev. {\bf B  14}, 3 (1976).
\bibitem{Mineevz2} V.P. Mineev, M. Sigrist, Phys. Rev. {\bf B  63}, 172504 (2001).
\bibitem{Rudnei} R.O. Ramos. D.G. Barci and C.A. Linhares, Brazilian Journal of Physics {\bf  37}, 1B (2007).
\bibitem{Wang} Jing Wang  and Guo-Zhu Liu, Phys. Rev.  D {\bf 90}, 125015 (2014).
\bibitem{Meissner}  K.  A.  Meissner and H. Nicolai, Acta Phys. Polonica B {\bf 40}, 2737 (2009).
\bibitem{Weinberg} E.J. Weinberg and A. Wu, Phys. Rev. D, {\bf 36}, 2474 (1987).
\bibitem{mermin-wagner} N.D. Mermin and H. Wagner, Phys. Rev. Lett., {\bf 17}, 1133 (1966).
\end{thebibliography}
\end{document}